\documentclass[pra,twocolumn,aps,floatfix,showpacs,tightenlines,showpacs,superscriptaddress]{revtex4}
\usepackage{amssymb,amsmath,amsthm,amsbsy,epsfig,color,graphicx}

\newcommand{\beq}{\begin{eqnarray}}
\newcommand{\eeq}{\end{eqnarray}}
\newcommand{\be}{\begin{equation}}
\newcommand{\ee}{\end{equation}}

\newcommand{\adots}{\reflectbox{$\ddots$}}

\begin{document}
\title{Genuine Multipartite Entanglement in the Cluster-Ising Model}

\author{S. M. Giampaolo}
\affiliation{University of Vienna, Faculty of Physics, Boltzmanngasse 5, 1090 Vienna, Austria}

\author{B. C. Hiesmayr}
\affiliation{University of Vienna, Faculty of Physics, Boltzmanngasse 5, 1090 Vienna, Austria}

%\date{\today}

%%%%%%%%%%%%%%%%%%%%%%%%%%%%%%%%%%%%%%%%%%%%%%%%%%%%%%%%%%%%%%%%%%%%%%%%%%%%%%%

\begin{abstract}
We evaluate and analyze the exact value of a measure for local genuine tripartite entanglement in the one-dimensional cluster-Ising model. This model is attractive since cluster states are considered to be relevant sources for applying quantum algorithms and the Ising interaction is an expected perturbation. Whereas bipartite entanglement is identically vanishing, we find that genuine tripartite entanglement is non zero in the anti-ferromagnetic phase and also in the cluster phase well before the critical point. We prove that the measure of local genuine tripartite entanglement captures all the properties of the topological phase transition. Remarkably, we find that the amount of genuine tripartite entanglement is independent of whether the considered ground states satisfy or break the symmetries of the Hamiltonian. We provide also strong evidences that for this experimentally feasible model local genuine tripartite entanglement represents the unique non vanishing genuine multipartite entanglement among any spins.
\end{abstract}

%{Quantum entanglement, scaling phenomena in complex systems, quantum phase transitions}

\maketitle

Quantum many-body systems usually posses a highly entangled ground state that exhibits collective quantum phenomena. Therefore, the analysis of the entanglement properties of a ground state becomes an important resource that flanks the standard analysis based on the Grinzburg-Landau paradigm~\cite{Amico2008} that is based on spontaneous symmetry breaking and a non-vanishing order parameter, e.g., ``magnetization''~\cite{Barber}. Although the presence and the importance of entanglement is recognized in many different physical contexts and physical systems, entanglement detection and quantification is a highly involved task (for mixed states). Except for the easiest case, bipartite spin-$\frac{1}{2}$ systems, no necessary and sufficient general method is known. The problem becomes much more pronounced for multipartite entanglement which has been found to play an essential role in many phenomena and may provide a high-capacity resource for quantum computing. Among multipartite entanglement \textit{genuine} multipartite entanglement is the most interesting one since all subsystems truly contribute to the entanglement. For example genuine multipartite entanglement is shown to be the necessary property for protecting a secret shared among many parties against eavesdropping or unfaithfully parties~\cite{SHH}.

In this letter we relate two rich phenomena in modern physics, i.e. the quantum phenomenon entanglement and (topological) phase transitions. In condensed matter systems phase transitions are at the heart of our understanding of critical phenomena. Previous works have shown that phase transitions can be characterized in terms of bipartite entanglement~\cite{Osterloh2002,Osborne2002} or in terms of the von Neumann entropy of bipartite spin-block
entanglement~\cite{Vidal2003,Latorre2004,Chen2010}, via global geometric entanglement~\cite{Wei2005} or via analyzing the existence
of points in which the entanglement is completely absent~\cite{factorization1,factorization2,factorization3}. Obviously, such methods fail for condensed matter systems that do not exhibit bipartite entanglement. Recent works~\cite{Giampaolo2013,Hofmann2013} have shown the presence of genuine multipartite entanglement in subsystems of typical condensed matter systems by computing criteria capable to detect different types of genuine multipartite entanglement.

This contribution shows that the exact amount of genuine tripartite entanglement can be computed exploiting a measure for genuine multipartite entanglement and that this measure characterizes fully the phase transition. In particular, our model exhibits a topological phase; those phases are known to play a prominent role e.g. in the quantum Hall effect~\cite{Kane}, for superconductivity~\cite{Volovik}, for the confinement problem in QCD or in string theory. Surprisingly, we find for all ground states, those preserving and those breaking the symmetries of the Hamiltonian, the same amount of genuine tripartite entanglement. The condensed matter model that we investigate has in one limit a pure Ising interaction and in the other limit  a cluster state as a ground state. Cluster states are a special type of multi-qubit states of graph states which are conjectured to be important resources for quantum algorithms (see, e.g., Ref.\cite{Bruss2}). This cluster-Ising model can be put to reality, e.g., for cold atoms in a triangular optical lattice~\cite{Pachos2004,Becker2010}. In the thermodynamic limit the anti-ferromagnetic phase is characterized by a standard staggered local order parameter whereas in the cluster phase one finds a gapped energy spectrum~\cite{Popp2005} with diverging localizable entanglement due to a non-vanishing string order parameter~\cite{CamposVenuti2005}.  In summary, this system is interesting since it can be experimentally put to reality, it is fully analytically solvable, has a quantum phase transition from an anti-ferromagnetic phase to a symmetry protected topological phase~\cite{Kou2009,Chen2010,Son2011,Smacchia2011} and provides a physical platform for quantum computation~\cite{Briegel2001,Miyake2010,Renes2011,Bruss2014}.

\begin{figure}[t]
\includegraphics[width=7.5cm]{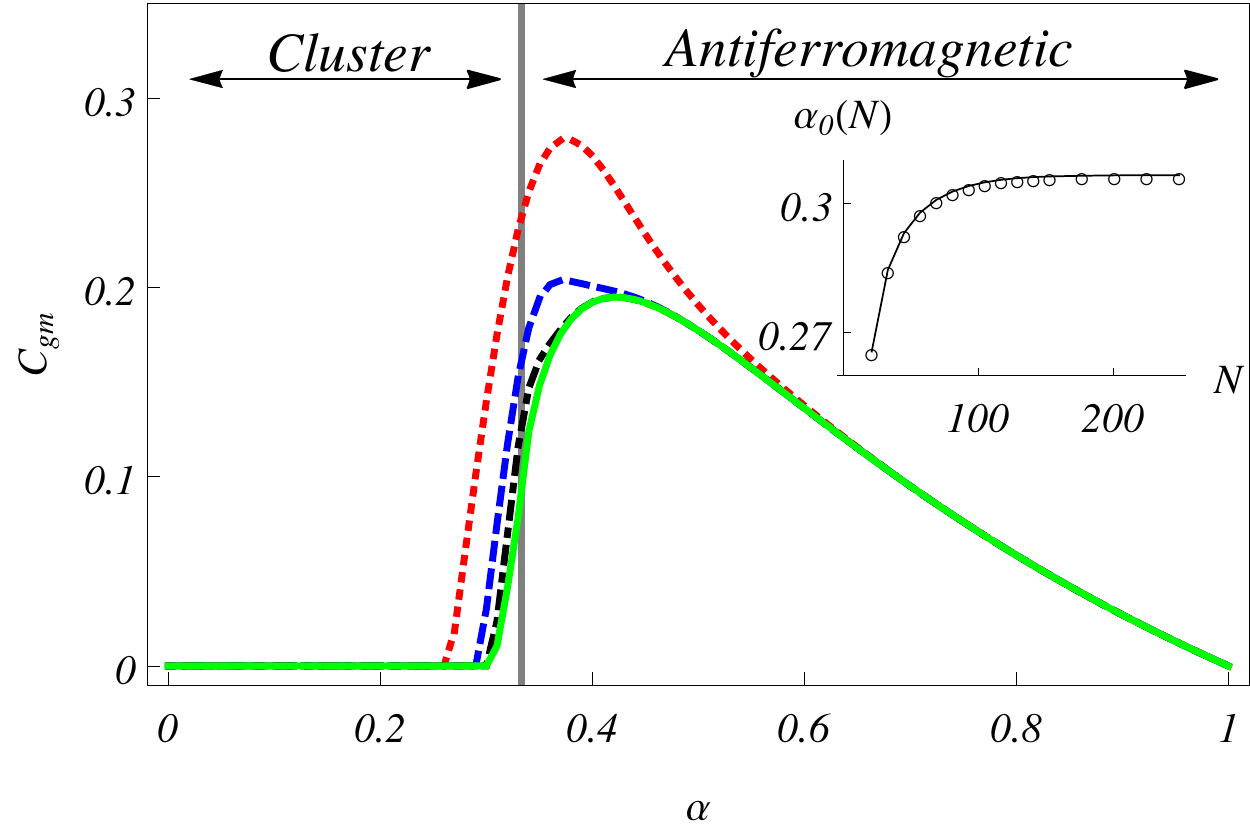}
\caption{(Color online) The genuine tripartite concurrence $\mathcal{C}_{gm}$ plotted for the three central spins in the chain as function of the
parameter $\alpha$ for different lengths $N$ of the spin chain. The different curves correspond to:
red dotted line $N=21$; blue dashed line $N=45$; black dot-dashed line $N=93$; green solid line $N=\infty$. The vertical grid line
indicate the critical value $\alpha_c=1/3$. In the inset we plot
the behavior of $\alpha_0(N)$, i.e. the value $\alpha$ for which the measures becomes nonzero, as function of the the length of the chain.}
\label{plotGMCfuncalpha}
\end{figure}

We start our analysis by introducing the Hamiltonian and the definition of genuine multipartite entanglement. We find that the reduced density matrix of three adjacent spins obtained from the symmetry preserving
ground state of the cluster-Ising model holds the property that it can be turn always in the $X$-form~\cite{Yu2007}. The $X$-form allows to compute analytically the exact amount of the \textit{genuine multipartite entanglement}~\cite{mconcurrence1,mconcurrence2,Hashemi2012} which is in this case identically to the general criteria for detecting different types of multipartite entanglement introduced in Refs.~\cite{Biseparabilitypaper1,Biseparabilitypaper2}. We then prove that a non vanishing value of such a measure of genuine tripartite entanglement characterizes as well the anti-ferromagnetic phase as the cluster phase close to the quantum phase transition (except for the factorization point~\cite{factorization1,factorization2,factorization3}). In particular, we prove that this measure captures all critical properties known as the universality property, i.e. the behaviour at the critical point is independent of the finite size scaling. In a second step we extend our analysis to ground states breaking the $Z_2$ symmetry of the Hamiltonian in the anti-ferromagnetic phase. We find that even in this case the reduced density matrix posses the same value of genuine tripartite entanglement, i.e. symmetry breaking does not affect the amount of entanglement. Last but not least we consider different subsets and analyze them according to their genuine multipartite entanglement content. Our results suggest that the genuine tripartite entanglement between three adjacent spins is the only source of genuine multipartite entanglement in this model.

To set the stage for our analysis let us introduce the one dimension cluster-Ising model, that is characterized by the
interplay between a three-body cluster like interactions and a two-body anti-ferromagnetic Ising term
\begin{equation}
 \label{hamiltonian}
 H=(-1+\alpha) \sum_{i=-l}^{l} S_{i-1}^x S_{i}^z S_{i+1}^x+\alpha  \sum_{i=-l-1}^{l} S_i^y S_{i+1}^y \;.
\end{equation}
The number of spins in the chain are $N=2l+3$ and %(in all our numerical simulation $N$ is a integer multiple of 3)
$\alpha$ is the relative weight between the two different interactions and may assume any values in the interval $[0,1]$.
$S_i^\mu$ are the standard spin-$1/2$ operators acting on the $i$-th site. We point out that all the evaluation made
for finite size are done assuming open boundary conditions and $N$ equal to an integer multiple of $3$.

Regardless its apparent complexity the cluster-Ising model can be analytically solved making use of the Jordan-Wigner transformation
that brings spin operators in fermionic ones~\cite{Jordan1928}. With this transformation the spin model is mapped into a bilinear fermionic
problem that can be analytically solved both for a finite size system and in thermodynamic limit (see appendix~\ref{suppmaterial}). Thanks to the existence of such analytical
solutions, the phase diagram of the one dimensional cluster-Ising model in the thermodynamic limit can be computed~\cite{CamposVenuti2005,Smacchia2011}. For
$\alpha>\frac{1}{3}$ the system is in an anti-ferromagnetic phase characterized by a twofold degenerate ground state with a non
vanishing staggered local order parameter along the $y$ axis $m_y=(-1)^i\langle\sigma_i^y \rangle$. On the contrary when
$\alpha<\frac{1}{3}$  the system is in the so called {\em cluster phase} with a four fold degenerate ground state that is characterized by
a highly non local string order parameter.

\begin{figure}[t]
\includegraphics[width=7.5cm]{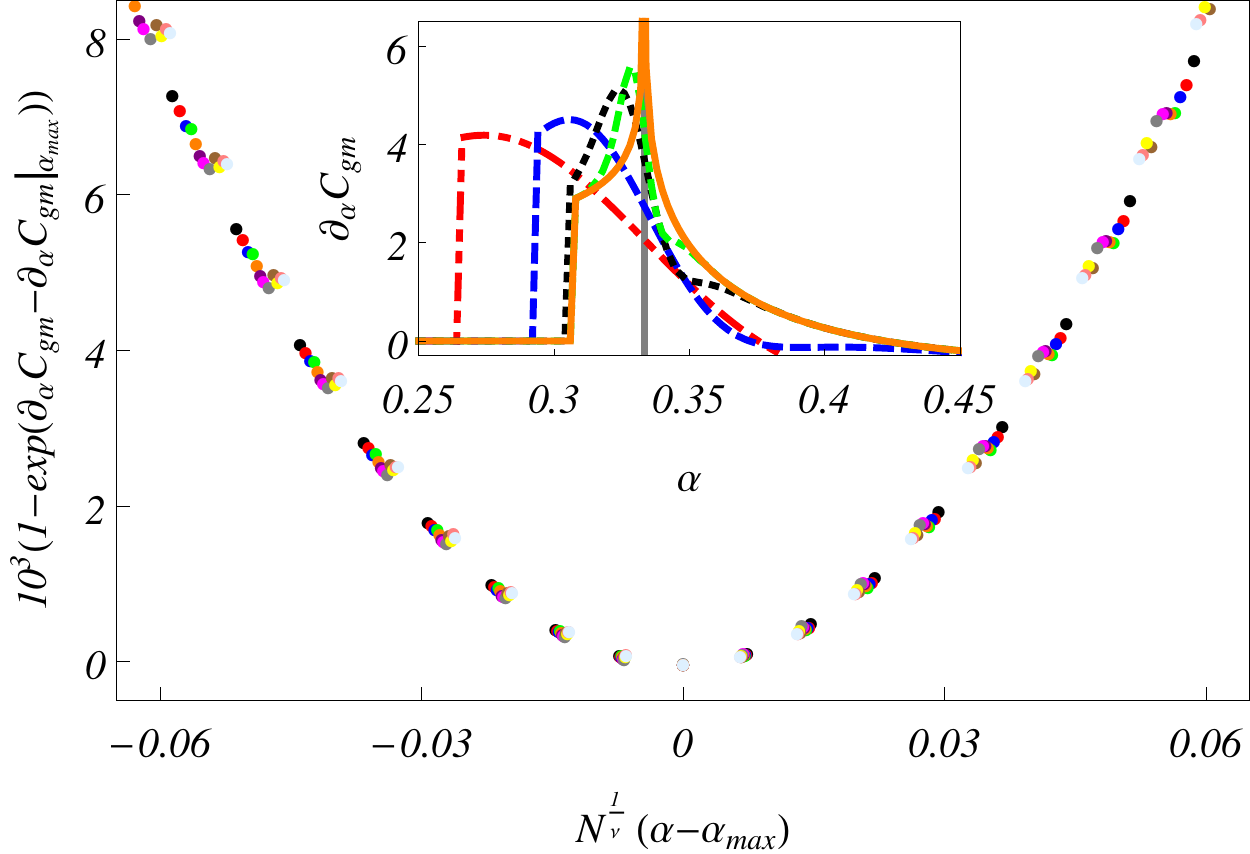}
\caption{(Color online) The universality (i.e., the fact that critical properties depend only on the size of the system and the broken symmetry in
the ordered phase) of the (rescaled) genuine multipartite entanglement is checked by plotting the finite-size scaling with respect to
$\mathcal{C}_{gm}$, according to the procedure described in Ref.~\cite{Barber}. The different colors represent different data that goes from $45$ to $249$ spins.
The critical exponent is taken equal to $\nu=1$. In the inset we plot
the behavior of $\partial_\alpha \mathcal{C}_{gm}$ close to $\alpha_c$ for chains of different lengths as function of $\alpha$: red
dot-dot-dashed line $N=21$; blue dashed line $N=45$; black dotted line $N=129$; green dot-dashed line $N=249$, orange solid line $N=\infty$.}
\label{plotgmcmaxderalpha}
\end{figure}

The first remarkable result concerns the local entanglement properties in the
cluster-Ising chain~\cite{Smacchia2011}. For any two spins taken out of the chain the entanglement is zero for all $\alpha$ (e.g., computed via Hill-Wootters concurrence~\cite{Hill1997,Wootters1998} which is a necessary and sufficient measure of entanglement of bipartite spin $\frac{1}{2}$ systems). Clearly, the vanishing of any pairwise entanglement disables the characterizing of the phase transition or the critical properties of the complex system. This result, together with the fact that the residual tangle is always fixed to its maximum value for the ground state that satisfies the symmetry of the Hamiltonian and never drops to zero also for the symmetry broken states~\cite{Smacchia2011,coffman2000,osburne2006} (except for $\alpha=1$), suggest that, in the
cluster-Ising model an important role must be played by the genuine multipartite entanglement.

Let us now define \textit{genuine} multipartite entanglement. A state is called $k$-separable if and only if it can be written in the form
\begin{eqnarray}
 \label{k-sep}
 \rho_{(k-sep)}&=&\sum_i p_i |\Psi_{(k-sep)}^i\rangle\langle \Psi_{(k-sep)}^i|
\end{eqnarray}
with $p_i\geq 0$ and $\sum_i p_i=1$ and $
|\Psi_{(k-sep)}\rangle= |\psi_1\rangle \otimes |\psi_2\rangle \otimes \ldots |\psi_k\rangle
$,
where $k\leq n$ and each $\psi_i$ ($i=1,\dots k$) is living on a different and non-overlapped Hilbert subspace. Note that biseparable states do not need to be separable via a specific partition since one sums up over different configurations. States which are not biseparable ($k=2$) are dubbed genuine multipartite entangled.

Computing the reduced density for three spins with equal distances $r$, $\rho_{i-r,i,r+i}$ ($r=1,2,\dots$), we found that for all values of $\alpha$ these reduced density matrices are in the so called $X$-matrix form~\cite{Yu2007}, i.e. only diagonal or off-diagonal elements are non-zero:

\begin{equation}
 \label{xform}
 \rho=\left(\begin{array}{cccccccc}
        t_1 & & & & & & & z_1 \\
            & t_2 & & & & & z_2 &  \\
         & & \ddots & & & \adots & &  \\
         & &  & t_n & z_n&  & &  \\
         & &  & z_n^* & s_n &  & &  \\
         & & \adots & & & \ddots & &  \\
         & z_2^* & & & & & s_2 &  \\
        z_1^* & & & & & & & s_1 \\
      \end{array}\right) \; ,
\end{equation}
In such a case it is possible to evaluate the genuine multipartite concurrence ($\mathcal{C}_{gm}$)~\cite{mconcurrence1,mconcurrence2} that is
a faithful measure of the genuine $m$-partite tangle ($m\!=\!1\!+\!\log_2 n$). In the presence of a density matrix in $X$ form the
$\mathcal{C}_{gm}$ is given by~\cite{Hashemi2012}
\begin{equation}
 \label{definitionGMC2}
 \mathcal{C}_{gm}\;=\;2 \max \biggl\lbrace0,|z_i|-\sum_{j\neq i} \sqrt{s_j t_j}\biggr\rbrace\;.
\end{equation}

Let us first consider the state $\rho_3^T$, a subset of three central spins $\{-1, 0, 1\}$ from the thermal ground state that preserves all the symmetries of the Hamiltonian,
\begin{eqnarray}
  \label{rho3t}
  \rho_3^T & = &\frac{1}{8} \left[ 1\!\!1- a\; (\sigma_{-1}^y \sigma_0^y +\sigma_{0}^y \sigma_1^y)+a^2\; \sigma_{-1}^y \sigma_1^y
  -b\; \sigma_{-1}^x \sigma_0^z  \sigma_1^x \right. \nonumber \\
  & + & \left. b a\; (\sigma_{-1}^x \sigma_0^x  \sigma_1^z+\sigma_{-1}^z \sigma_0^x  \sigma_1^x)+b a^2\; \sigma_{-1}^z \sigma_0^z  \sigma_1^z \right]
\end{eqnarray}
where $a\!=\!-G(1,\alpha)\!=\!\langle \sigma_0^y \sigma_1^y \rangle$, $b\!=\!G(-2,\alpha)\!=\!-\langle \sigma_{-1}^x \sigma_0^z \sigma_1^x \rangle$ and
$G(n,\alpha)$ are the fermionic correlation functions (see appendix~\ref{suppmaterial}). In the case of a finite size system
 the fermionic correlation functions are evaluated numerically using an algorithm that generalize the approach of Ref.~\cite{Lieb1961}. On the contrary in the thermodynamic limit they are given by~\cite{Smacchia2011}
\begin{equation}
 \label{Gfunction}
 G(n,\alpha)=\frac{1}{\pi} \int_{0}^\pi \frac{\cos(n+2)x-\frac{2\alpha}{1-\alpha} \cos(n-1)x}{\Lambda(x,\alpha)} dx
\end{equation}
where
\begin{equation}
 \label{Lambda}
 \Lambda(x,\alpha)=\left[1+\left(\frac{2 \alpha}{1-\alpha}\right)^2 -\frac{4 \alpha}{1-\alpha} \cos(3x)\right]^{\frac{1}{2}}\;.
\end{equation}

Regardless the value of the fermionic correlation functions, applying the unitary operator
\begin{equation}
 \label{unitary}
 U=e^{-i \frac{\pi}{4} \sigma_{-1}^x} \otimes e^{-i \frac{\pi}{4} \sigma_{0}^x} \otimes e^{-i \frac{\pi}{4} \sigma_{1}^x}
\end{equation}
turns $\rho_3^T$ into the $X$-form with the additional symmetry that $t_i=s_i$ with $i=1,\dots,n$ (corresponding to the vanishing of
$\langle \sigma_i^\gamma \rangle=0 \;\; \forall \gamma=x,y,z$). From Eq.~(\ref{rho3t}) it is straightforward to obtain the value
of the genuine tripartite entanglement as
\begin{equation}
 \label{valueGMC}
 \mathcal{C}_{gm}= \frac{1}{4} \max \left\{0,(1+a)^2(1+b)-4\right\}\;,
\end{equation}
which behavior as function of $\alpha$ is plotted  both in thermodynamic limit and for finite size systems in Fig.~\ref{plotGMCfuncalpha}. Independently, of the number of spins $N$ the behavior of the measure $\mathcal{C}_{gm}$ is similar and converges with increasing number fast to a nonzero value for the anti-ferromagnetic phase (except the factorization point $\alpha=1$). Remarkably, it becomes non-zero already well before the critical point $\alpha_c=\frac{1}{3}$ and also independently of $N$. In the inset of  Fig.~\ref{plotGMCfuncalpha} we have plotted the dependence of the value of $\alpha$ when it becomes non-zero in dependence of $N$. For the thermodynamic limit we obtain $\alpha_0(\infty) \simeq 0.3064$, a value clearly below $\alpha_c=1/3$. This rising of the genuine tripartite measure before the quantum phase transition plays the roles of a precursor of the quantum phase transition.

We can go one step further by considering the derivative $\partial_\alpha\mathcal{C}_{gm}=\partial \mathcal{C}_{gm}/\partial \alpha$ and exploring its non trivial behavior close to the quantum critical point. For the finite size case we find a maximum before the critical point that converges with increasing number of sizes $N$ to a maximum at the critical point (see inset of Fig.~\ref{plotgmcmaxderalpha}). Being more precise the maximum value of
$\partial_\alpha\mathcal{C}_{gm}$ diverges logarithmically with $N$ as
\begin{equation}
 \label{finitesizescaling1}
 \left. \partial_\alpha \mathcal{C}_{gm} \right|_{\alpha=\alpha_{max}}=0.64 \log N + const\;,
\end{equation}
and in the thermodynamic limit the derivative $ \partial_\alpha \mathcal{C}_{gm}$ diverges approaching the critical value
$\alpha=\alpha_c\equiv \frac{1}{3}$:
\begin{equation}
 \label{finitesizescaling2}
  \lim_{N\longrightarrow\infty}\;\partial_\alpha \mathcal{C}_{gm}=0.64 (- \log |\alpha-\alpha_c|) + const\;.
\end{equation}
According to the scaling ansatz~\cite{Barber,Osterloh2002,Giampaolo2013}, which validity is tested in Fig.~\ref{plotgmcmaxderalpha}, in the
case of logarithmic singularities, the ratio between the two prefactors of the logarithm in Eq.~(\ref{finitesizescaling1}) and
Eq.~(\ref{finitesizescaling2}) is the exponent $\nu$ that governs the divergence of the correlation length
$\xi\approx|\alpha-\alpha_c|^{-\nu}$. We find $\nu=1$ that is consistent with the characterization obtained from the analysis of the behavior
of the correlation function.

In the cluster phase, regardless the huge degeneracy of the ground space, there is no ground state that breaks the symmetry of the
Hamiltonian, in contrast to the anti-ferromagnetic phase. Let us superpose two ground states that each preserves in the anti-ferromagnetic phase all the symmetries of the Hamiltonian, then we obtain a maximally symmetry broken ground state which reduced density
matrix of three adjacent spins is given by
\begin{equation}
  \label{rho3b}
  \rho_3^B = \rho_3^T+ \frac{1}{8} \left[ c \, (\sigma_{-1}^y -\sigma_0^y +\sigma_1^y) -d\, \sigma_{-1}^y \sigma_0^y \sigma_1^y \right]
\end{equation}
where $c$ and $d$ are the expectation values of $\langle \sigma_i^y \rangle_B $ and
$\langle \sigma_{i-1}^y \sigma_{i}^y \sigma_{i+1}^y \rangle_B$, respectively. The subscript $B$ indicates that the average is evaluated on
states breaking maximally the symmetry. Since the additional terms are only proportional to $\sigma^y$ the $X$-form can be recovered using the same
local unitary transformation as defined in Eq.~(\ref{unitary}). Moreover, as shown in the supplementary materials, $c=d$ holds and, consequently, the same value of the genuine tripartite measure is obtained. This also holds for any ground state that breaks the $Z_2$ symmetry. Hence, we conclude that $\mathcal{C}_{gm}$ does not depend on the choice of the ground state and, therefore, captures all the relevant properties of the system.

\begin{figure}[t]
\includegraphics[width=7.cm]{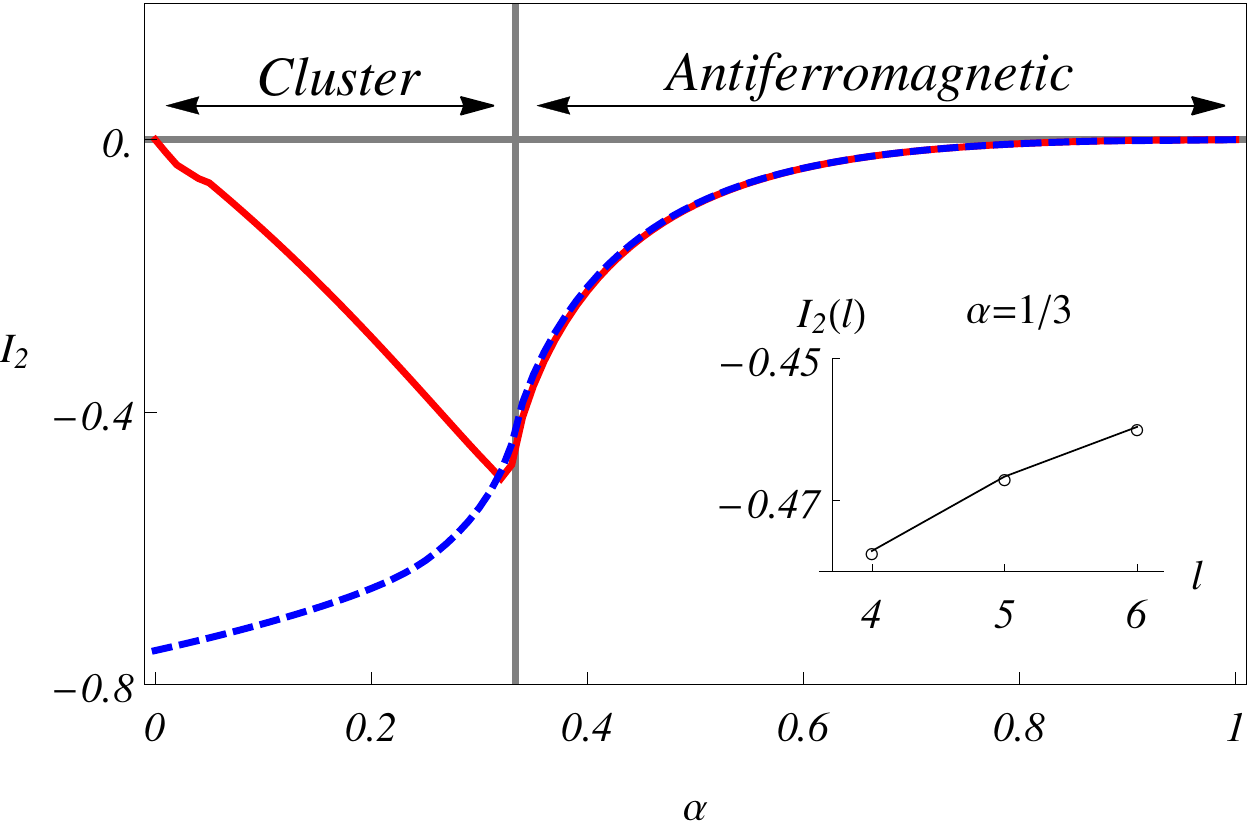}
\caption{(Color online) The function $I_2$ detecting genuine multipartite entanglement~\cite{Biseparabilitypaper1,Biseparabilitypaper2} if greater than zero is plotted as function of $\alpha$ for
the subsets of three spins $\{-2, 0, 1\}$ (blue dashed lines) and the subsets of four spins $\{-2, -1, 0, 1\}$ (red solid lines). The inset shows $I_2$ for $l=4,5,6$ adjacent spins for $\alpha=\frac{1}{3}$. No genuine multipartite entanglement is detected.}
\label{plotI2}
\end{figure}

Let us now investigate the entanglement properties of different subsets. Increasing the distance between the
spins but still preserving the symmetry of the subset, i.e. considering the subset $\{-s, 0, s\}$ with $s=2,3,\dots$ we find that the reduced density matrices can be brought into the $X$-form but the measure vanishes. If we increase the subset, e.g. considering the four spin subset $\{-2, -1, 0, 1\}$ the reduced density matrix can no longer be brought into the $X$-form. This is due to the presence of non vanishing two body correlation function along the $x$ direction $\langle \sigma_i^x \sigma_j^x \rangle$.

In Fig.~\ref{plotI2} we plotted the function $I_2$ introduced in Refs.~\cite{Biseparabilitypaper1,Biseparabilitypaper2}, which is a lower bound on genuine multipartite entanglement, for three and four adjacent spins. The heart of this approach are convex functions of the density matrix $\rho$ and permutations operators $\mathcal{P}$ on subsets $\beta$ defined on copies of $\rho$ which are symmetries of a certain type of multipartite entanglement, i.e. $[\mathcal{P}_\beta,\rho^{\otimes 2}]=0\Longrightarrow \rho^{\otimes 2}=\mathcal{P}_\beta\rho^{\otimes 2}\mathcal{P}_\beta$. If the symmetry is satisfied the following convex function is bounded by
\begin{eqnarray}
 \label{criterion}
 &&I_k\!:=\!\\
 &&|\langle\Phi_1|\rho|\Phi_2\rangle|-\sum_{\{\beta\}}\!\left(\! \prod_{j=1}^k
 \langle\Phi_1 \Phi_2|P_{\beta,j}^\dagger\rho^{\otimes 2}\! P_{\beta,j}|\Phi_1\Phi_2\rangle\!\right)^{\frac{1}{2k}}\le \; 0\nonumber
\end{eqnarray}
for $\Phi_{1/2}$ some arbitrary states with dimension of $\rho$ and $j$ denotes a certain partition in the subset $\beta$ and $k$ defines the number of partitions one is interested in. The states satisfying the above inequality for a given partition $k$ are just the $k$-separable states defined in Eq.~(\ref{k-sep}), hence, any positive values detects states that are not $k$-separable.

Fig.~\ref{plotI2} shows that no genuine multipartite entanglement is detected for any value of $\alpha$, moreover, the value is rather far away from the bound zero. In the inset of Fig.~\ref{plotI2} we plotted the value of $I_2$ at the critical point  for four to six adjacent spins. We observe only a very small in
crease of the value. These results strongly suggest that there is no other type of genuine multipartite entanglement present than the genuine tripartite one of three adjacent spins.

Summarizing, in the present letter we have provided the first evaluation of the genuine multipartite entanglement in the ground state of
a complex quantum systems for which bipartite entanglement is zero. We have determined analytically the genuine tripartite entanglement for a subset made by three adjacent spins in dependence on weight $\alpha$ between the cluster and the anti-ferromagnetic interactions in a finite size system and in the thermodynamic limit. We find for any size of the system that local genuine tripartite entanglement is non-zero before the critical point $\alpha=\frac{1}{3}$, i.e. genuine tripartite entanglement acts as a precursor of the phase transition. Moreover, we have shown that the non-trivial behaviour of the genuine tripartite entanglement measure characterizes the critical and scaling properties fully. We have provided strong results that the genuine tripartite entanglement seems to be the only manifestation of genuine multipartite entanglement in this particular condensed matter system. In future perspective, it would be desirable to extend such analysis to other condensed matter systems and to try to connect the presence of genuine multipartite entanglement both with the behavior of the
entanglement spectrum~\cite{Giampaolo2013_2} and with scaling property of the quantum frustration~\cite{Giampaolo2011,Marzolino2013}.

{\em Acknowledgments} - SMG and BCH acknowledge gratefully the Austrian Science Fund (FWF-P23627-N16) and fruitful discussions with M. Huber and SMG with O. G\"uhne.

\newpage

\section{Appendix}\label{suppmaterial}

In this section we describe how the Hamiltonian is diagonalized, subsequently how one obtains the fermionic correlation functions and from them the spin correlations functions that are analyzed in the main manuscript.

\subsection{Diagonalization of the Hamiltonian}

Despite the presence of three-body interactions the Hamiltonian~(\ref{hamiltonian}) can be mapped into a delocalized fermionic form
via a generalization of the approach described in Ref.~\cite{Lieb1961}. To obtain this map we introduce the Jordan-Wigner transformation~\cite{Jordan1928} that connects the spin operators to the fermionic
operator via
\begin{equation}
 \label{JW_transformation}
 c_j=\left(\bigotimes_{i=-l-1}^{j-1} 2 S_i^z \right) S_j^- \;, \, \, \, \, \, \,
 c_j^\dagger=\left( \bigotimes_{i=-l-1}^{j-1} 2 S_i^z \right) S_j^+\, ,
\end{equation}
where $S_i^\pm=S_i^x\pm i S_i^y$. Applying this transformation to the Hamiltonian~(\ref{hamiltonian}) we obtain
% \begin{eqnarray}
%  \label{hamiltonian_2}
%  H\!\!&\!=\!&\!\! \frac{-1+\alpha}{8}\!\!\sum_{i=-l}^{l}\! c_{i-1}^\dagger c_{i+1}^\dagger\! -\! c_{i-1} c_{i+1}\!+\! c_{i-1}^\dagger c_{i+1}\!+
%  \!c_{i+1}^\dagger c_{i-1} \nonumber \\
%  & +&\frac{\alpha}{4} \sum_{i=-l-1}^{l} c_{i}^\dagger c_{i+1}^\dagger - c_{i} c_{i+1}+ c_{i}^\dagger c_{i+1}+c_{i+1}^\dagger c_{i}
% \end{eqnarray}
\begin{eqnarray}
 \label{hamiltonian_2}
 H&=& \frac{1-\alpha}{8} \sum_{i=-l}^{l} (c_{i-1}^\dagger c_{i+1}^\dagger + c_{i-1}^\dagger c_{i+1} + h.c.)\nonumber \\
 & +&\frac{\alpha}{4} \sum_{i=-l-1}^{l} (c_{i}^\dagger c_{i+1}^\dagger +c_{i}^\dagger c_{i+1} + h.c)\;.
\end{eqnarray}
This  Hamiltonian corresponds to a so-called free fermion case in which non-interacting fermions can jump between nearest and next to nearest neighbors of the chain via open boundary conditions, i.e. $N=2l+3$.
This Hamiltonian can be brought in a diagonal form
\begin{equation}
 \label{hamiltonian_3}
 H= \sum_k \lambda_k\; \eta_k^\dagger \eta_k + const.
\end{equation}
using the following linear transformation
\begin{eqnarray}
\label{eta}
\eta_k&=& \sum_{i=-l-1}^{l+1} \frac{\phi_{k,i} + \psi_{k,i}}{2} c_i+  \frac{\phi_{k,i} - \psi_{k,i}}{2} c_i^\dagger \nonumber \\
\eta_k^\dagger&=& \sum_{i=-l-1}^{l+1} \frac{\phi_{k,i} + \psi_{k,i}}{2} c_i^\dagger+  \frac{\phi_{k,i} - \psi_{k,i}}{2} c_i \, ,
\end{eqnarray}
where both vectors $\phi_k$ and $\psi_k$ are a set of $N$ vectors with $N$ components that are solutions of the coupled equations
\begin{eqnarray}
\label{psipsh1}
\lambda_k \psi_k&=& \phi_k  (\mathbf{A}+\mathbf{B})\nonumber \\
\lambda_k \phi_k&=& \psi_k (\mathbf{A}-\mathbf{B})  \,.
\end{eqnarray}
Here the matrix $\mathbf{A}$ (symmetric) and $\mathbf{B}$ (anti-symmetric) are, respectively,
\begin{equation}
\label{matA}
\mathbf{A}\!=\!\frac{1}{8}\!\left(\!
\begin{array}{cccccccccc}
 0\!\!&2 \alpha\!\!& \gamma \!\!&0\!\!&0\!\!&0\!\!&0\!\!&0\!\!&0\!\!&0 \\
 2 \alpha\!\!& 0 \!\!& 2 \alpha\!& \gamma\!&0\!&0\!&0\!&0\!&0\!&0 \\
 \gamma\!\!& 2\alpha \!& 0 \!& 2 \alpha\!& \gamma\!&0\!&0 \!&0\!&0\!&0 \\
 \!\!& \ddots\!& \ddots\!& \ddots\!& \ddots\!& \ddots \!& \\
\!\!& \!& \ddots\!& \ddots\!& \ddots\!& \ddots\!& \ddots \!& \\
\!\!& \!& \!& \ddots\!& \ddots\!& \ddots\!& \ddots\!& \ddots \!& \\
\!\!& \!& \!& \!& \ddots\!& \ddots\!& \ddots\!& \ddots\!& \ddots \!& \\
 0\!\!&0 \!&0\!&0\!&0\!&\gamma\!& 2\alpha \!& 0 \!& 2 \alpha\!& \gamma\\
 0\!\!& 0\!&0 \!&0\!&0\!&0\!&\gamma\!& 2\alpha \!& 0 \!& 2 \alpha \\
 0\!\!&0\!& 0\!&0 \!&0\!&0\!&0\!&\gamma\!& 2\alpha \!& 0
\end{array}
\!\right)
\end{equation}
and
\begin{equation}
\label{matB}
\mathbf{B}\!=\!\frac{1}{8}\!\left(\!\!
\begin{array}{cccccccccc}
\!\! 0\!&2 \alpha\!&\! \gamma \!&\!0\!&\!0\!&\!0\!&\!0\!&\!0\!&\!0\!&\!0 \\
\!\! -2 \alpha\!&\! 0 \!&\! 2 \alpha\!&\! \gamma\!&\!0\!&\!0\!&\!0\!&\!0\!&\!0\!&\!0 \\
\!\! -\gamma\!&\! -2\alpha \!&\! 0 \!&\! 2 \alpha\!&\! \gamma\!&\!0\!&\!0 \!&\!0\!&\!0\!&\!0 \\
\!\! \!&\! \ddots\!&\! \ddots\!&\! \ddots\!&\! \ddots\!&\! \ddots \!&\! \\
\!\!\!&\! \!&\! \ddots\!&\! \ddots\!&\! \ddots\!&\! \ddots\!&\! \ddots \!&\! \\
\!\!\!&\! \!&\! \!&\! \ddots\!&\! \ddots\!&\! \ddots\!&\! \ddots\!&\! \ddots \!&\! \\
\!\!\!&\! \!&\! \!&\! \!&\! \ddots\!&\! \ddots\!&\! \ddots\!&\! \ddots\!&\! \ddots \!&\! \\
\!\! 0\!&\!0 \!&\!0\!&\!0\!&\!0\!&\!-\gamma\!&\! -2\alpha \!&\! 0 \!&\! 2 \alpha\!&\! \gamma\\
\!\! 0\!&\! 0\!&\!0 \!&\!0\!&\!0\!&\!0\!&\!-\gamma\!&\! -2\alpha \!&\! 0 \!&\! 2 \alpha \\
\!\! 0\!&\!0\!&\! 0\!&\!0 \!&\!0\!&\!0\!&\!0\!&\!-\gamma\!&\! -2\alpha \!&\! 0
\end{array}
\!\!\right)
\end{equation}
and $\gamma=1-\alpha$.
Either $\psi_k$ or $\phi_k$ can be eliminated from the equations~(\ref{psipsh1}) resulting in
\begin{eqnarray}
\label{psipsh2}
\lambda_k^2 \psi_k&=& \psi_k (\mathbf{A}+\mathbf{B})(\mathbf{A}-\mathbf{B})\nonumber \\
\lambda_k^2 \phi_k&=& \phi_k (\mathbf{A}-\mathbf{B})(\mathbf{A}+\mathbf{B})  \,.
\end{eqnarray}
For $\lambda_k \neq 0$ we can choose one of these two equations that gives us one set of vectors and with the help of the euations.~(\ref{psipsh1}) one obtains the second set of vectors. For $\lambda_k = 0$ one obtains $\psi_k$ and $\phi_k$ from the equations~(\ref{psipsh1}).

\subsection{Fermionic correlation functions}

The set of vectors $\psi_k$ and $\phi_k$ plays an important role for the evaluation of the fermionic correlations functions.
To see that let us introduce two new fermionic operators
\begin{equation}
\label{otherfermionic}
 \nu _i= c^\dagger_i + c_i \,; \,\,\,\,\,\,\,\,\,\,\,\,\,\, \mu_i = c^\dagger_i - c_i \, .
\end{equation}
%All the spin correlation functions can be transformed in fermionic correlation function that we can evaluate easily making use of the
%Wick's Theorem~\cite{Wick} that allow us to express the expectation value of a product of fermionic operators in terms of the so-called
%contractions of pairs, i.e. expectation value of product of just two operators.
The advantage to use this set of operators, $\mu_i$ and
$\nu_i$, with respect to the set of operators $c_i$ and $c_i^\dagger$ comes from the fact that, taking into account the Eqs.~(\ref{eta}),
we obtain particular simplifications computing the expectation values of the product of two operators, in particular:
\begin{eqnarray}
\label{correlations}
 \langle \mu_i \mu_k\rangle = & \sum_l \phi_{l,i} \phi_{l,j} = & \delta_{ik} \nonumber \\
 \langle \nu_i \nu_k\rangle = & - \sum_l \psi_{l,i} \psi_{l,j} = & -\delta_{ik}  \\
 \langle \nu_i \mu_k\rangle = -\langle \mu_i \nu_k\rangle = & - \sum_l \psi_{l,i} \phi_{l,j} = & G_{i,j}(\alpha)\;. \nonumber
\end{eqnarray}
In the finite size case the $G_{i,j}(\alpha)$ depends on the choice of the site $i$ and $j$ and, therefore, also the spin correlation functions
depend on the position of the spins in the chain. However, if we evaluate the correlation functions between three central spins of
a chains with an odd $N$, the mirror symmetry plays the role of the invariance under spatial translation, imposing that
\begin{equation}
\langle S_{-1}^\alpha S_{0}^\beta \rangle=\langle S_{1}^\alpha S_{0}^\beta \rangle \quad \forall\; \alpha,\beta=x,y,z\;.
\end{equation}

On the contrary, in the thermodynamic limit the invariance under spatial translation is restored by the fact that the end of the chain is
infinitely far away. In this case, imposing the invariance under spatial translation, we can solve Eqs.~(\ref{psipsh1}) considering that
each components of $\phi_k$ and  $\psi_k$ vectors obey to the following rule
\begin{equation}
\phi_{k,l}=\frac{q_{k,l}}{\sqrt{N}}e^{-i \frac{2 \pi}{N} k l} \, \, ; \, \, \, \, \, \psi_{k,l}=\frac{q'_{k,l}}{\sqrt{N}}e^{-i \frac{2 \pi}{N} k l} \, .
\end{equation}
With a long but straightforward evaluation we obtain the fermionic correlation functions in the thermodynamic limit given in Eq.~(\ref{Gfunction}) and Eq.~(\ref{Lambda}), respectively.
%\begin{equation}
% \label{Gfunction_supp}
% G(n,\alpha)=\frac{1}{\pi} \int_{0}^\pi \frac{\cos(n+2)x-\frac{2\alpha}{1-\alpha} \cos(n-1)x}{\Lambda(x,\alpha)} dx
%\end{equation}
%where $n=2 \pi k /N$ and $\Lambda(x,\alpha)$ is equal to
%\begin{equation}
% \label{Lambda_supp}
% \Lambda(x,\alpha)=\left[1+\left(\frac{2 \alpha}{1-\alpha}\right)^2 -\frac{4 \alpha}{1-\alpha} \cos(3x)\right]^{\frac{1}{2}}
%\end{equation}
%From the expression of Eq.~(\ref{Gfunction_supp}) we may immediately seen that if $n$ cannot be written as $n=3m+1$ with $m$ integer, than
%the fermionic correlation function is equal to zero. The same property also holds for the finite chain when the length $N$ is an integer
%multiple of 3.

\subsection{Spin correlation functions}

Having introduced a method to evaluate the fermionic correlation functions we can compute the spin correlation functions from which we obtain any
reduced density matrix.
We can divide the spin correlation functions in two families. The first is consists of correlation functions from operators that do not break the
symmetry of the Hamiltonian. In this case transforming the spin operators into the fermionic operators $\mu_i$ and $\nu_i$ one always obtains
an operator constructed via an equal number of $\mu_i$ and $\nu_i$. Applying the Wick's theorem we then obtain the expression of the correlation
functions in terms of the $G(n,\alpha)$. Let us also point out that the dependence of the spin correlation functions on the fermionic
correlation functions are independent of the model. A list of these spin correlation functions given in terms of fermionic correlation functions
can be founded in Ref.~\cite{Hofmann2013}.

The situation is much more involved in the case the symmetry of the
Hamiltonian is broken. In this second case, if we try to apply the same strategy as described above, we immediately observe that
the elements due to parity immediately vanish. In fact, re-writing the operators in terms of $\mu_i$ and $\nu_i$, we obtain an
expression with different numbers of fermionic operators, each one referring to distinct sites, and, in agreement with Eq.~(\ref{correlations}),
this implies that the correlation function must be zero. This result holds for all ground states that preserve the symmetry of the
Hamiltonian, however, in the thermodynamic limit and in the anti-ferromagnetic phase one has also ground states that break the $Z_2$
symmetry of the Hamiltonian. For these ground states some of these correlation functions are expected to be different from zero.

\begin{figure}[t]
{\includegraphics[width = 7cm]{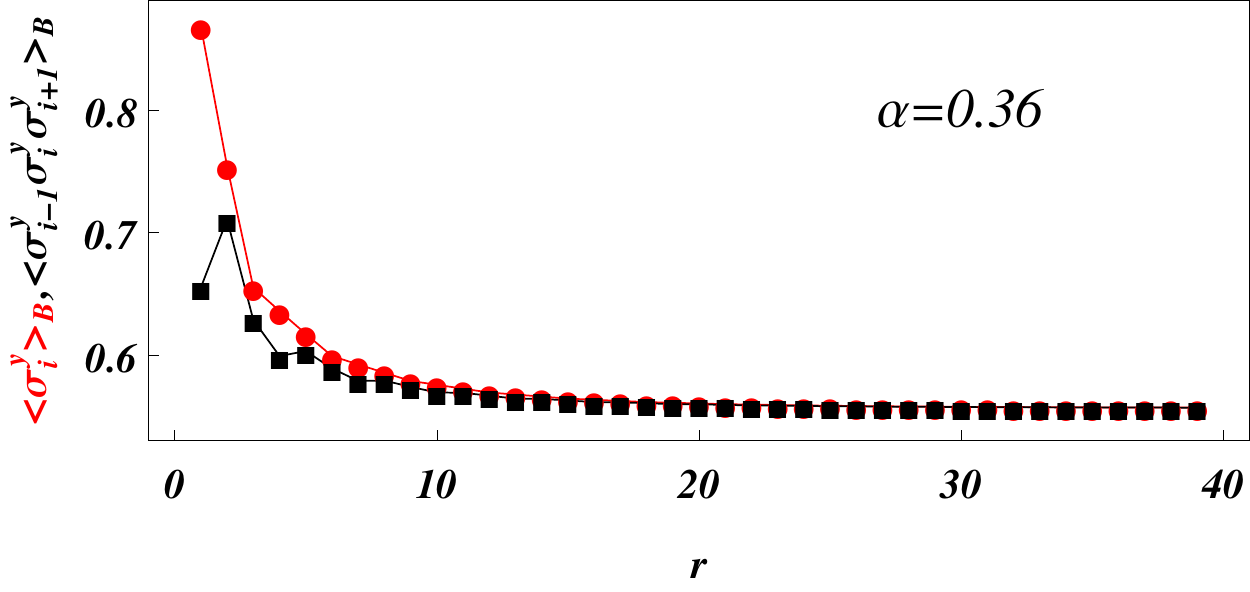}}\\
{\includegraphics[width = 7cm]{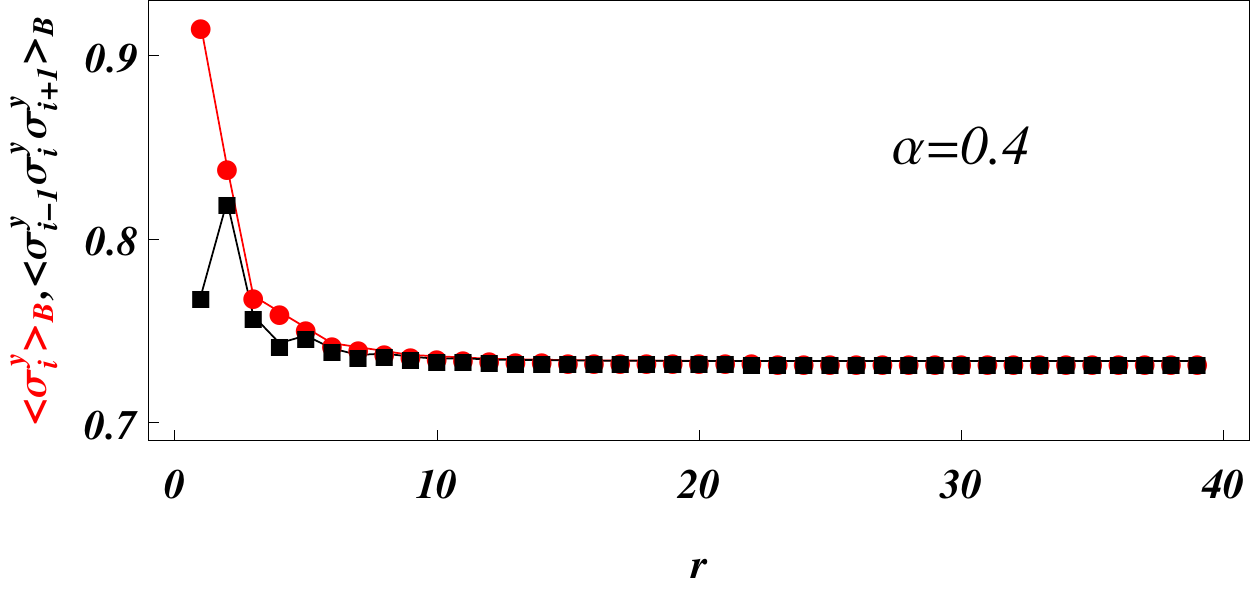}}\\
{\includegraphics[width = 7cm]{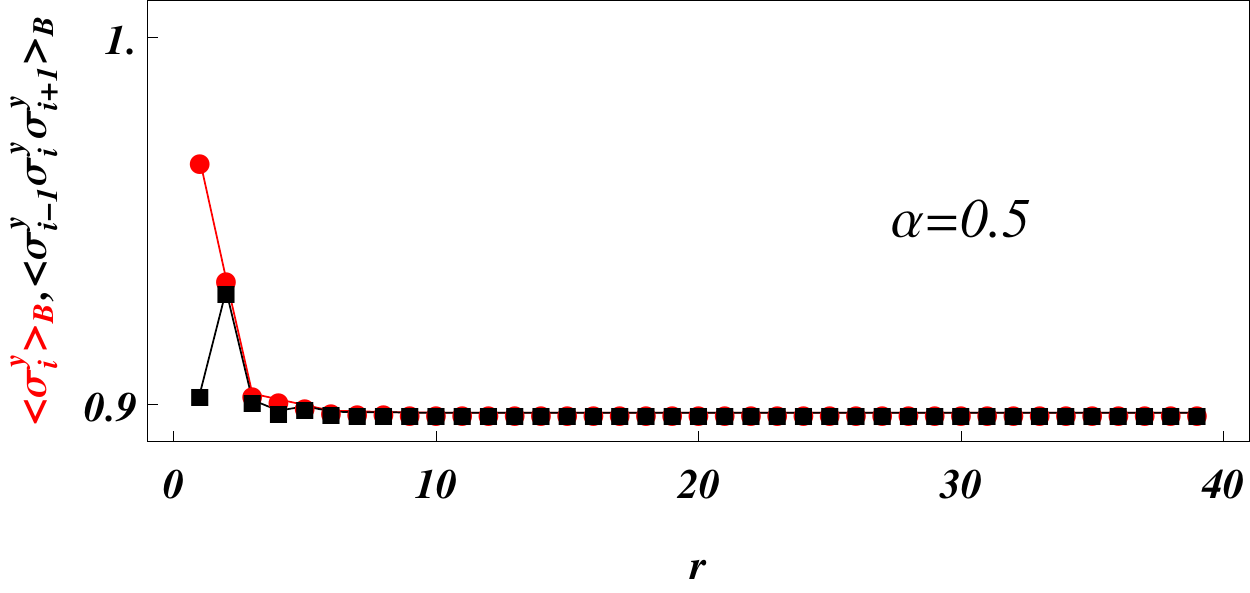}}
\caption{(Color online) The figures show the values of $\langle \sigma_i^y \rangle_B $ (red points) and
$\langle \sigma_{i-1}^y \sigma_{i}^y \sigma_{i+1}^y \rangle_B$ (black squares) for different distances $r$ and for
different values of $\alpha$ in the anti-ferromagnetic phase.}
\label{plotsupp1}
\end{figure}

To evaluate these correlation functions we adopt the approach of Barouch and McCoy~\cite{BarouchMcCoy1971}. Given an operator
$O_{\{k\}}$ defined on a set of spins $\{k\}$ that does not commute with the parity operator along the $Z$ axis one can define the correlation function
$\langle O_{\{k\}}\rangle_B$ via
\begin{equation}
 \label{corr-break}
 \langle O_{\{k\}} \rangle_B=\lim_{r \rightarrow \infty} \sqrt{\langle O_{\{k\}} O_{\{k\}+r}\rangle}\;.
\end{equation}

For analyzing the entanglement properties of three adjacent spins one
has to compute all the one, two and three body correlation functions that do not commute with the parity operators along the $Z$ axis. All
the correlation functions of this type are vanishing in the limit of $r \rightarrow \infty$ except $\langle \sigma_i^y \rangle_B $ and
$\langle \sigma_{i-1}^y \sigma_{i}^y \sigma_{i+1}^y \rangle_B$. In Fig.~\ref{plotsupp1} we have reported, for some values of $\alpha$, the values
of  $\langle \sigma_i^y \rangle_B $ and $\langle \sigma_{i-1}^y \sigma_{i}^y \sigma_{i+1}^y \rangle_B$. We observe that the farer we are away from the quantum critical point the faster is the convergence of the correlation functions to the asymptotic values. Moreover, for all $\alpha$ we observe that $\langle \sigma_i^y \rangle_B = \langle \sigma_{i-1}^y \sigma_{i}^y \sigma_{i+1}^y \rangle_B$ for large enough $r$
(we have tested this relations for more that $1000$ different value of $\alpha>\alpha_c$). The fact that this equality holds impose that the
genuine tripartite entanglement between three adjacent spins does not depends on the particular ground state selected as emphasized in the main body of the manuscript.

\end{document}